# Analyzing the Data of COVID-19 with Quasi-Distribution Fitting Based on Piecewise B-spline Curves

Qingliang Zhao[2]; Zhenhuan. Lu[2]; Yiduo. Wang[1,a*]
[1]Department of Mathematics, Beijing University of Chemical Technology, China.
[2]The College of Economics and Management, Beijing University of Chemical Technology, China.
[a]victor_wang@yeah.net, *Corresponding author

**Abstract**
Facing the world wide coronavirus disease 2019 (COVID-19) pandemic, a new fitting method (QDF, quasi-distribution fitting) which could be used to analyze the data of COVID-19 is developed based on piecewise quasi-uniform B-spline curves. For any given country or district, it simulates the distribution histogram data which is made from the daily confirmed cases (or the other data including daily recovery cases and daily fatality cases) of the COVID-19 with piecewise quasi-uniform B-spline curves. Being dealt with area normalization method, the fitting curves could be regarded as a kind of probability density function (PDF), its mathematical expectation and the variance could be used to analyze the situation of the coronavirus pandemic. Numerical experiments based on the data of certain countries have indicated that the QDF method demonstrate the intrinsic characteristics of COVID-19 data of the given country or distric, and because of the interval of data used in this paper is over one year (500 days), it reveals the fact that after multi-wave transmission of the coronavirus, the case fatality rate has declined obviously, the result shows that as an appraisal method, it is effective and feasible.

## 1  Introduction

In late December 2019, cases of pneumonia with unknow aetiology were reported in the city of Wuhan, China[1]. The causative agent, identified as the betacoronavirus SARS-CoV-2, is closely related to SARS-CoV which was responsible[2] for the outbreak of SARS between 2002 and 2004. SARS-CoV-2 caused a sizable epidemic of COVID-19 in China, then spread globally and declared a pandemic in March 2020[3]. Until now, there have been a lot of research articles about this pandemic, demonstrated the plague in different angles. Some compared its pathogenesis with that of previously Middle East respiratory syndrome (MERS) and SARS[4], or indicated how COVID-19 pneumonia compromise the distal lung performs essential respiratory functions[5], or detailed virological analysis of cases of COVID-19 that provides proof of active virus replication in tissues of the upper respiratory tract[6]. Some discussed the COVID-19-related mortality among different ages[7], gender[8], or race[9], shed light on the frequency of asymptomatic SARS-CoV-2 infection[10]. There were also some research works focus on building mathematics models to simulate the spread of SARS-CoV-2, to be specifically: metapopulation susceptible-exposed-infectious-removed (SEIR) model which integrated fine-grained, dynamic mobility networks simulating the spread of COVID-19 in ten of the largest US metropolitan areas[11], and full-spectrum dynamics model which reconstructed the transmission mode of COVID-19 in Wuhan between 1 January and 8 March 2020[12]. The analyzing of data showed anti-contagion polices have significantly and substantially slowed the growth of COVID-19 infections[13], and the major non-pharmaceutical interventions-and lockdowns in particular- have had a large effect on reducing transmission[14] in the same way. A research group studied relationship between socio-economic factors and the COVID-19 pandemic in Germany by analyzing both infections and fatalities, their results showed that the population of poorer and more socially deprived districts was not necessarily more likely to get infected with SARS-CoV-2, but combined an average infection rate with a higher than average death rate[15], beside these social problems caused by COVID-19, it also changed people's everyday life praxis during restriction measures[16].
Estimating the size of the coronavirus disease 2019 (COVID-19) pandemic is made challenging by inconsistencies in the available data[17], so it is very important to analyze the COVID-19 data of given country or district, inculding the number of daily confirmed cases, daily recovery cases and daily fatality cases. In this paper, we use a quasi-distribution-fitting method based on piecewise quasi-uniform B-spline curves to investigate the consistency of infection and the trending patterns across different countries. It is established by fitting the distribution histogram data made from the COVID-19 data of given country or district with piecewise quasi-uniform B-spline curves, dealt with area normalization process, the fitting curves could be regarded as a kind of probability density function (PDF) of the data, then calculate the mathematical expectation and the variance as evaluation result.

## 2 Theoretical considerations

In computer aided geometric design, the B-spline form is widely used in representing a polynomial curve. B-spline curves have optimal shape preserving properties, and a B-spline curve of order $n$ is evaluated by the de Casteljau algorithm with a computational cost of $O(n^2)$ elementary operations [18]. But B-spline curve has a shortcoming, it is a kind of curve whose control polygon is not combined with the curve itself at their endpoints, which means changing of even one control point, all the points in the curve will be changed, so, in this paper we will use piecewise quasi-uniform B-spline curve to fulfill the fitting works.

**Table 1 Daily new cases of different countries in five days**

| Finland | | France | | korea | | Ecuador | |
|---|---|---|---|---|---|---|---|
| 11/4/2020 | 293 | 7/2/2020 | 659 | 1/15/2021 | 513 | 7/13/2020 | 589 |
| 11/5/2020 | 189 | 7/3/2020 | 582 | 1/16/2021 | 1099 | 7/14/2020 | 0 |
| 11/6/2020 | 266 | 7/4/2020 | 0 | 1/17/2021 | 0 | 7/15/2020 | 1870 |
| 11/7/2020 | 0 | 7/5/2020 | 0 | 1/18/2021 | 389 | 7/16/2020 | 1036 |
| 11/8/2020 | 412 | 7/6/2020 | 1375 | 1/19/2021 | 386 | 7/17/2020 | 1079 |

### 2.1 Histogram distribution

Because of the delay in epidemiology statistic system of different countries or districts, we need to precondition the data before the fitting process. Table 1 shows five daily confirmed cases data of four countries, in the first column, we can figure out that daily confirmed cases in Finland during Nov 4, 2020 to Nov 6, 2020 is around 200 per day, but in Nov 7, 2020, it down to zero, and then back to 412 in Nov 8, 2020, almost twice of the previous data, the most probable explaination is the data of daily confirmed cases of Nov 7, 2020 was delayed, and then combined in the data of the following day. The same scenario happened in the data of France, Korea and Ecuador, the simply way to avoid it is to use adjacent average data instead of the original one. In this paper, we use 7-day moving average data to fulfill our research plan, which means in order to get the 7-day moving average data, we need extra three days' data out of both side of the data interval. For example, the data interval of Italy we used in this paper is from Feb 21, 2020 to July 4, 2021, containing 500 days' data, but the actual data we used is from Feb 18, 2020 to July 7, 2021. For given country or district, $D_k$ denote 7-day moving average of the daily confirmed cases number in day $k$, let $N$ be the data interval, in this paper $N=500$, hence, for Italy, $D_1$ is Feb 21's 7-day moving average of the daily confirmed cases which is the average of the daily confirmed cases from Feb 18, 2020 to Feb 24, 2020.

Assume $D_\sigma = \sum_{i=1}^{N} D_i$, let $f_k^D = \dfrac{D_k}{D_\sigma}$, then we have $\sum_{k=1}^{N} f_k^D = 1$

That means $f_k^D$ could be regarded as a kind of probability distribution, we can call that histogram distribution. The table 2 show the begin date and the end date of the countries whose data of COVID-19 would be used in our fitting process.

**Table 2 Beginning date and ending date of different countries' COVID-19 data**

| Country | Beginning date | Ending date | Country | Beginning date | Ending date |
|---|---|---|---|---|---|
| Austria | 3/01/2020 | 7/13/2021 | Brazil | 3/06/2020 | 7/18/2021 |

| Canada | 2/28/2020 | 7/11/2021 | Chile | 3/04/2020 | 7/16/2021 |
| --- | --- | --- | --- | --- | --- |
| US | 2/29/2020 | 7/12/2021 | Denmark | 3/02/2020 | 7/14/2021 |
| France | 2/26/2020 | 7/09/2021 | German | 2/29/2020 | 7/12/2021 |
| India | 3/02/2020 | 7/14/2021 | Iran | 2/19/2020 | 7/02/2021 |
| Israel | 3/06/2020 | 7/18/2021 | Italy | 2/21/2020 | 7/04/2021 |
| Japan | 1/29/2020 | 6/11/2021 | Korea | 2/17/2020 | 6/30/2021 |
| Lebanon | 2/26/2020 | 7/09/2021 | Philippine | 3/07/2020 | 7/19/2021 |
| Poland | 3/04/2020 | 7/16/2021 | Portugal | 2/24/2020 | 7/07/2021 |

## 2.2 Piecewise Quasi-uniform B-spline curve

For those histogram distribution data $f_k^D$, it could be simulated with a function $P(x)$, which has all the properties of probability density function (PDF), thus the histogram distribution data $f_k^D$ could be analyzed from the respective of probability theory. In this paper, we fit $f_k^D$ with piecewise quasi-uniform B-spline curve, which is more flexible in curve modeling. Piecewise quasi-uniform B-spine curve is a kind of parameter curve (here the parameter denote as $t \in [0,1]$), it is gotten from uniform B-spline curve. First, we give the base functions of quintic quasi-uniform B-spline curves defined on interval $[0,1]$, whose node vector is dividing $[0,1]$ into ten subintervals evenly, denote them as $\tilde{N}_i, i=0,1,\cdots 14$, the concrete expression of those base functions $\tilde{N}_i$ could be found in the appendix. Second, we define the base functions of piecewise quasi-uniform B-spline as:

$$N_i(t) = \begin{cases} \tilde{N}_i\left(\dfrac{t}{\omega}\right) & t \in [0,\omega) \\ 0 & t \in [\omega,1] \end{cases} \quad i=0,1,\cdots 13, \qquad N_{14}(t) = \begin{cases} \tilde{N}_{14}\left(\dfrac{t}{\omega}\right) & t \in [0,\omega) \\ \tilde{N}_0\left(\dfrac{1-t}{1-\omega}\right) & t \in [\omega,1] \end{cases},$$

$$N_i(t) = \begin{cases} 0 & t \in [0,\omega) \\ \tilde{N}_{i-14}\left(\dfrac{1-t}{1-\omega}\right) & t \in [\omega,1] \end{cases} \quad i=15,\cdots,28, \text{ where } \omega \in (0,1) \text{ is the segmentation point.}$$

Assume $C_i, i=0,1,\cdots 28$ are the points in two-dimensional plane, then the definition of quintic piecewise quasi-uniform B-spline curve is:

$$B(t) = \sum_{i=0}^{28} N_i(t) C_i \quad t \in [0,1]. \tag{1}$$

Where $C_i$ are called control points of the curve defined by equation (1).

## 3 Fitting process

For given data $f_k^D$, find a quintic piecewise quasi-uniform B-spline curve $B(t) = \sum_{i=0}^{28} N_i(t) C_i$ to fitting those data. The most important part is to calculate unknown control points $C_i$.

### 3.1 Least square approximation method

In this paper, we use the least square approximation to deal with this problem. First, parameterize those data $f_k^D$ by cumulative chord length parameterization method to match a parameter $t_k$ for every $f_k^D$, thus we get a parametric sequence $0 = t_1 < \cdots t_k < \cdots < t_N = 1$. Second, assume $\bar{P}_k = (k, f_k^D)$, $k=1,2,\cdots N$, then build a vector equation group which has $N$ equations to solve the unknown control points $C_i$:

$$B(t_k) = \sum_{i=0}^{28} N_i(t_k) C_i = \bar{P}_k, \quad k=1,2,\cdots N. \tag{2}$$

Equation (2) could be solved as:

$$\phi^T \phi \begin{bmatrix} C_0 \\ C_1 \\ \vdots \\ C_{28} \end{bmatrix} = \phi^T \begin{bmatrix} \bar{P}_0 \\ \bar{P}_1 \\ \vdots \\ \bar{P}_N \end{bmatrix}, \text{ where } \phi = \begin{bmatrix} N_0(t_1) & N_1(t_1) & \cdots & N_{28}(t_1) \\ N_0(t_2) & N_1(t_2) & \cdots & N_{28}(t_2) \\ \vdots & \vdots & & \vdots \\ N_0(t_N) & N_1(t_N) & \cdots & N_{28}(t_N) \end{bmatrix}. \quad (3)$$

In order to decide the segmentation point $\omega$, we need to figure out how to evaluate the goodness of a fitting result. Being a parameter fitting curve, $B(t)$ need to be discretized into a standard discrete signal $\bar{B}_k$ to match with $f_k^D$. Divide parameter interval $(0,1)$ into $n$ uniform parts ($n \gg N$), then get a discrete signal sequence $B_i = (B_i^x, B_i^y), i = 1, 2, \cdots n$. The x-coordinate of $\bar{P}_k$ is $k$, then we have: $\bar{B}_k = B_{k'}^y$, if $B_{k'}^x = \max_i \{B_i^x < k\}$ $k = 1, 2, \cdots N$. After that, we evaluate the goodness of the simulation result with mean square deviation (MSE), calculated as:

$$MSE(\bar{B}, f^D) = \frac{\sum_{i=1}^{N}(\bar{B}_i - f_i^D)^2}{N}. \quad (4)$$

Denote the standard discrete signal $\bar{B}$ which is gotten with the segmentation point $\omega'$ as $\bar{B}^{\omega'}$, then the best segmentation point $\omega$ could be decided like:

$$\omega = \omega' \Big|_{\omega' \in (0,1)} \min MSE(\bar{B}^{\omega'}, f^D) \quad (5)$$

### 3.2 Quasi-distribution

The fitting signal $\bar{B}_k$ is an approximation to the histogram distribution $f_k^D$, thus the sum of $\bar{B}_k$ may not be 1. Nevertheless, we can fulfill it through an adjustment factor $\gamma$, as:

$$\gamma = 1 \Big/ \sum_{k=1}^{N} \bar{B}_k \quad (6)$$

Let $\tilde{B}_k = \gamma \bar{B}_k$, then, $\tilde{B}_k$ satisfy the property of probability density function, but its expression is not like any existed probability density functions, we can call that quasi-distribution.

### 3.3 Experimental result

In this paper, we investigated eighteen countries' COVID-19 data with the data interval of 500 days, table 2 shows the beginning date and the ending date of each country, and the fitting of the quasi-distribution result show as the following figures. In those figures, the green colored signal denote 7-day moving average of the original data (called histogram data), including daily confirmed cases, daily recovery cases, and daily fatality cases, then with the algorithm just presented, we get the corresponding quasi-distribution fitting of those data, finally, we put those curves of quasi-distribution fitting together, to find the inner trend of the pandemic. Figure 1 shows the experimental result of Austria, in figure 1 (a), the red curve is the quasi-distribution fitting of the histogram data of daily confirmed cases, based on previous definition, it could be regarded as the probability density function of daily confirmed cases if it was assumed as random variable, in figure 1 (a), we can see, the quasi-distribution curve fitting the histogram data of daily conformed cases perfectly. Similarly, in figure 1 (b), the blue curve is the quasi-distribution fitting of the histogram data of daily recovery cases, in figure 1 (c), the black curve is the quasi-distribution simulation of the histogram data of daily fatality cases. In figure 1 (d), we put three fitting curves in the same coordinate, apparently they both have three peaks, but at the first peak, the fatality curve is higher than the other two, at the second peak, no obviously difference, but at the third peak, the fatality curve is quite lower than the other two, is that mean, with the time going on, even in the situation of the virus mutation, the case fatality rate of the COVID-19 is keep on declining? Or this situation only occurs in Austria, and the data of more countries need to be analyzed.

**Fig. 1.** Histogram data and quasi-distribution fitting of Austria

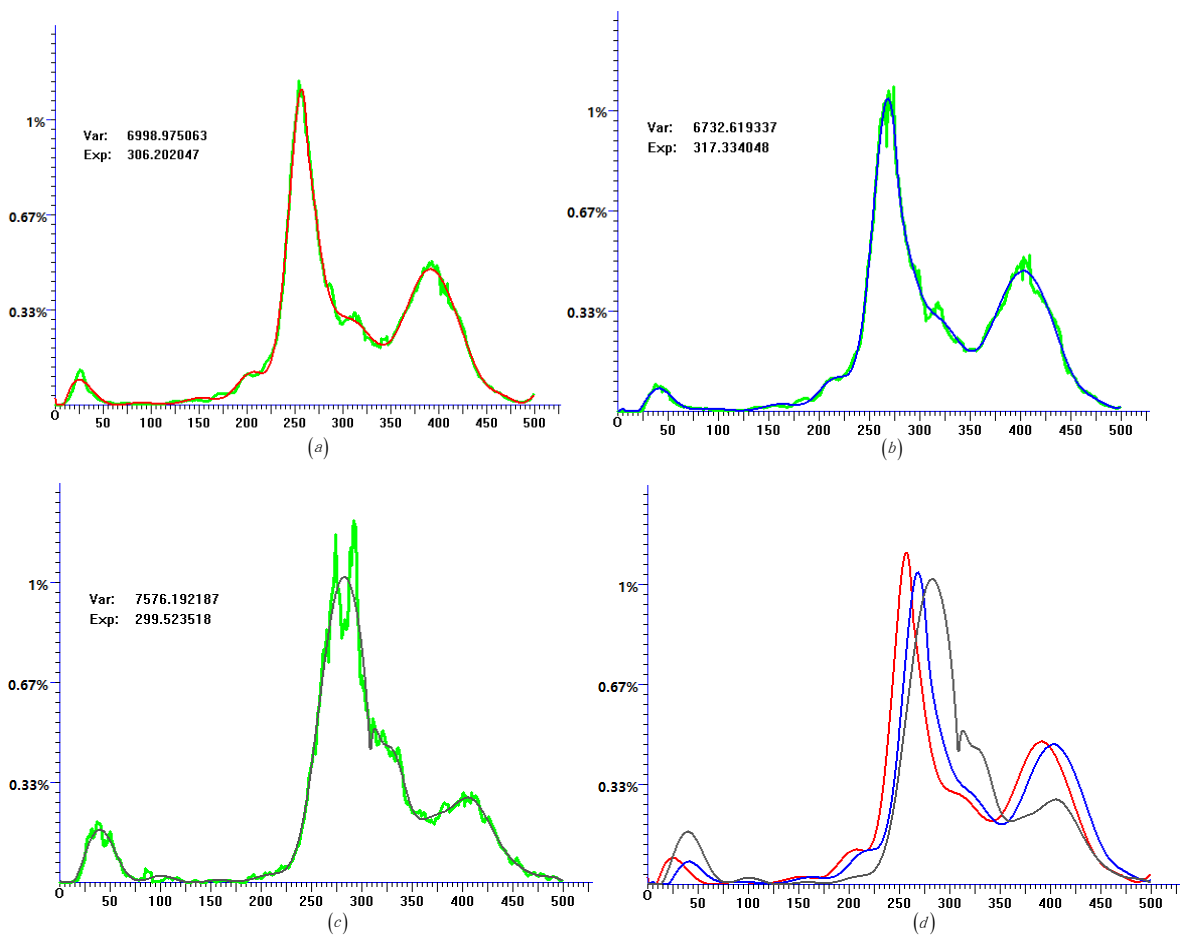

In figure 2, we made the same procedures on the data of Brazil, but the result seems not like that of Austria's, especially in figure 2 (d), the fatality peak around day 400, seems quite higher than the previous peak around day 100, but the trend of quasi-distribution fitting curves of daily confirmed cases and daily recovery cases are still quite similarly, so more countries' data need to be counted.

**Fig. 2.** Histogram data and quasi-distribution fitting of Brazil

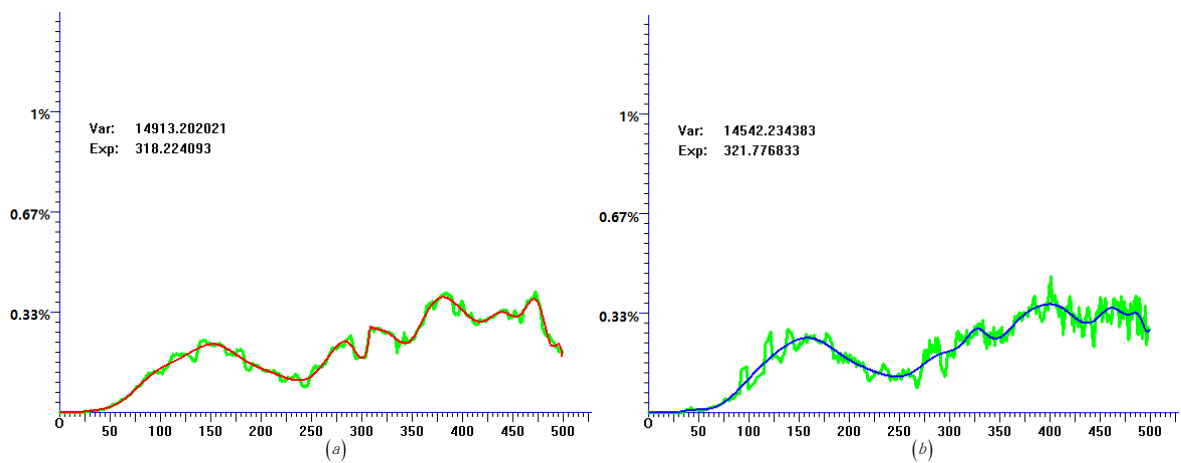

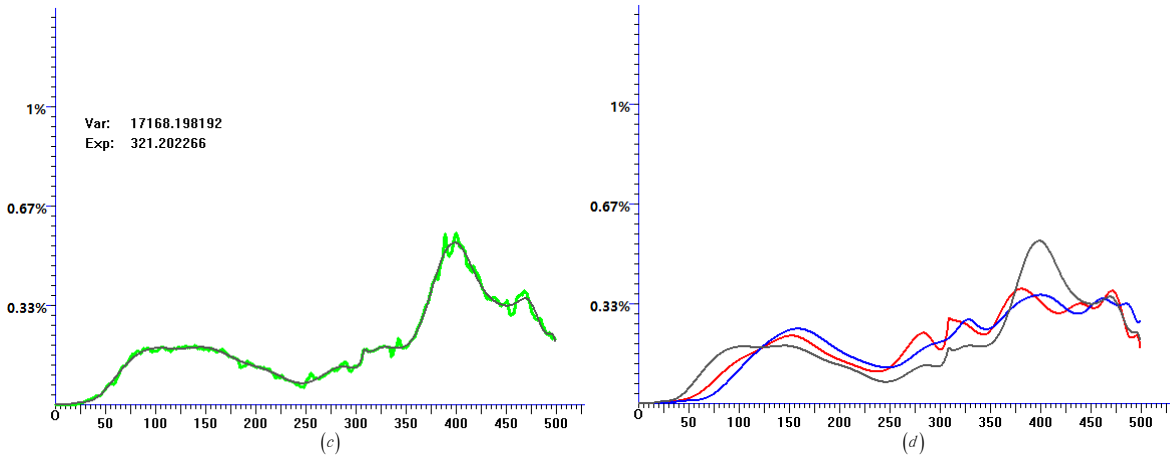

Figure 3 shows the experimental result of Canada, it shows the same result like Austria's, in figure 3 (d) the third peak of the fatality curve is much lower than that of the daily confirmed cases and daily recovery cases, and in figure 3 (b), around day 140 there is a booming of the daily recovery cases, apparently it was not because of the suddenly increasing of the medical system, but the releasing of accumulated data, that is why we do not using the histogram data (7-day moving average data) to discover the inner trend of the pandemic, but the corresponding quasi-distribution fitting curves.

**Fig. 3.** Histogram data and quasi-distribution fitting of Canada

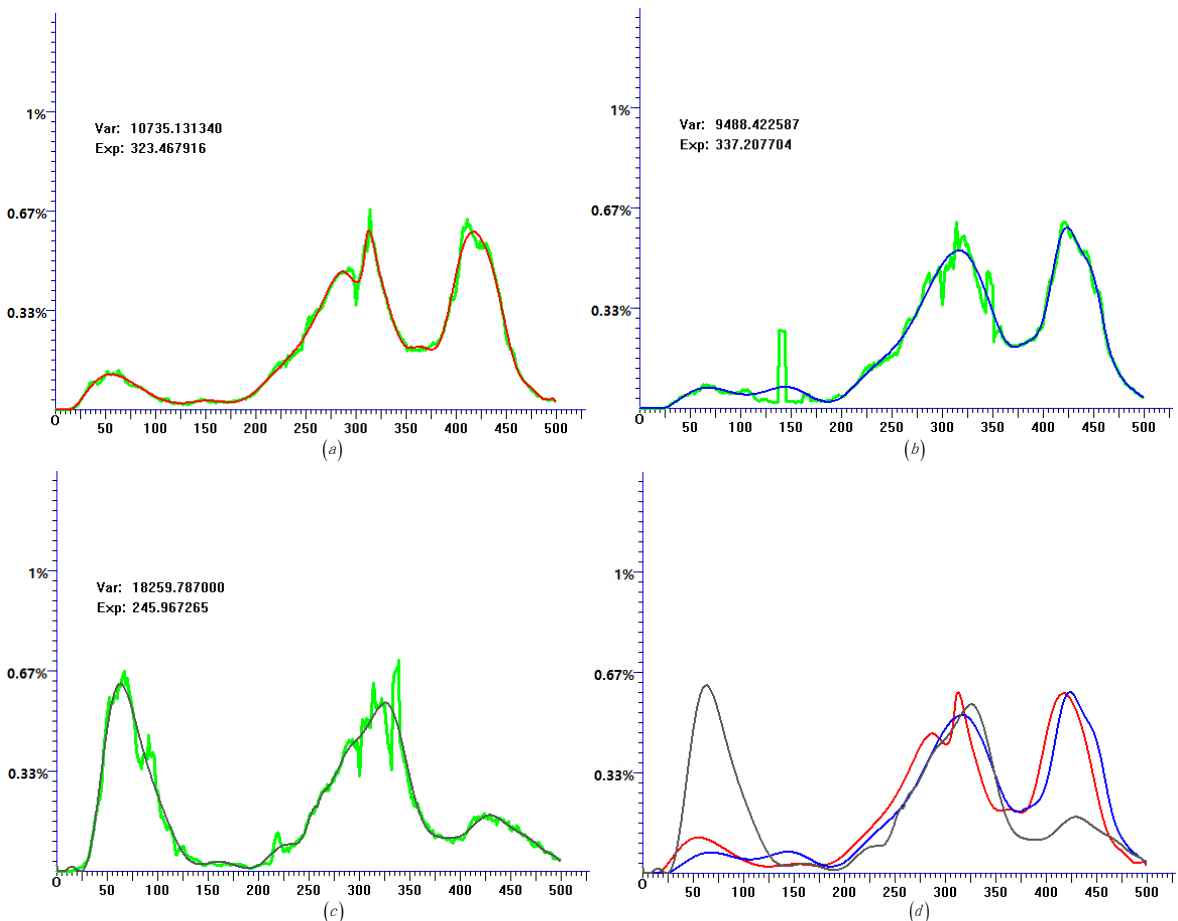

Figure 4 is the situation of Chile, demonstrate the same result of Austria and Canada, the peak of fatality curve in the later time are obviously lower than the peaks of daily confirmed cases and daily recovery cases, showing the trend of case fatality rate declining, and based on the value of "Var" (denote the variance of the quasi-distribution fitting curve) in (a), (b), (c), the pandemic mode of Chile is quite similar to Brazil rather than Austria and Canada, maybe it is because they are both located in South America and their geographical location is the same as the climatic environment, affecting the spread of the epidemic.

**Fig. 4.** Histogram data and quasi-distribution fitting of Chile

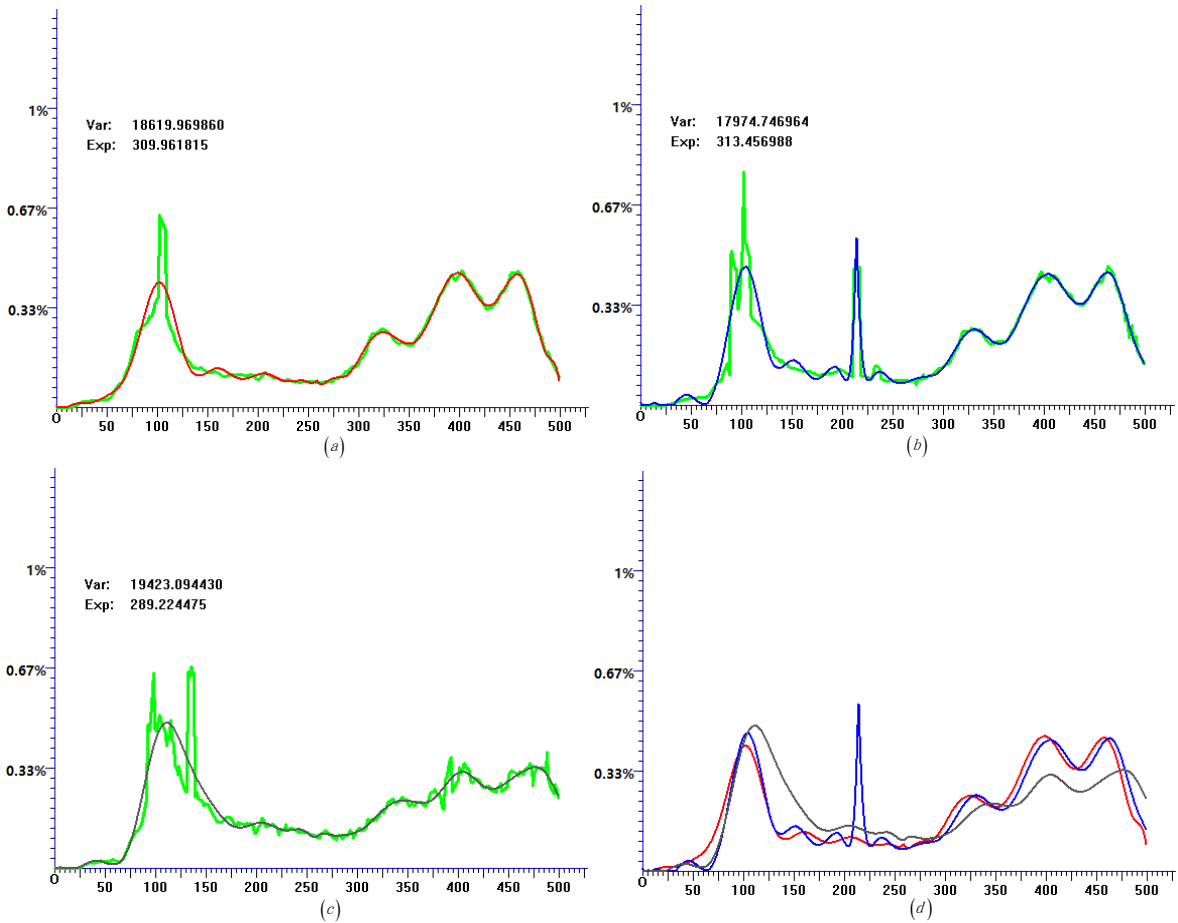

Figure 5，figure 6, figure 7, figure 8, figure 9, figure 10 show the fitting results of epidemic data in Denmark, France, German, Italy, Iran and US respectively, and they are almost the same of that in Austria and Canada, which demonstrated the fact that the third peak of quasi-distribution fitting curve made by daily fatality cases declined obviously. Accordingly, it means that with the spread of the epidemic, the fatality rate has decreased obviously.

**Fig. 5.** Histogram data and quasi-distribution fitting of Denmark

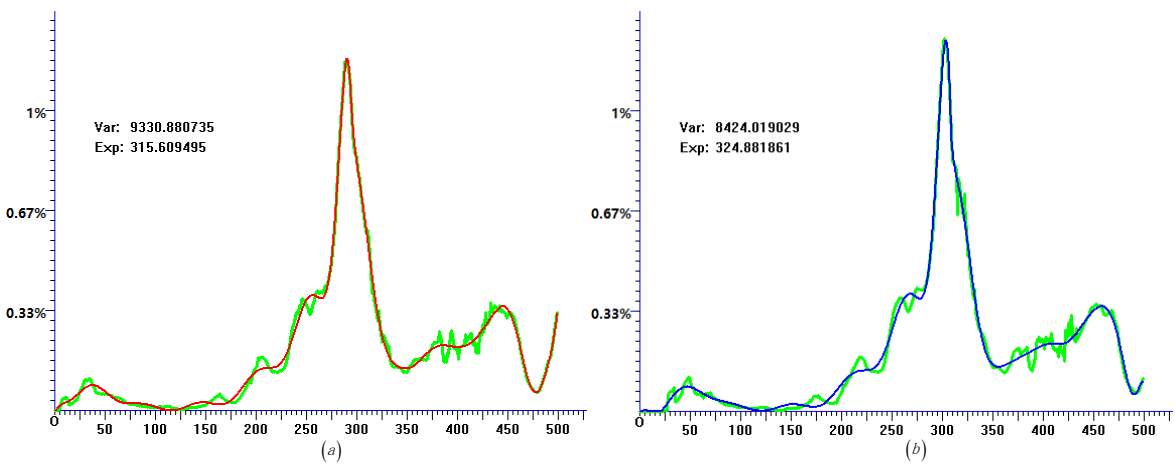

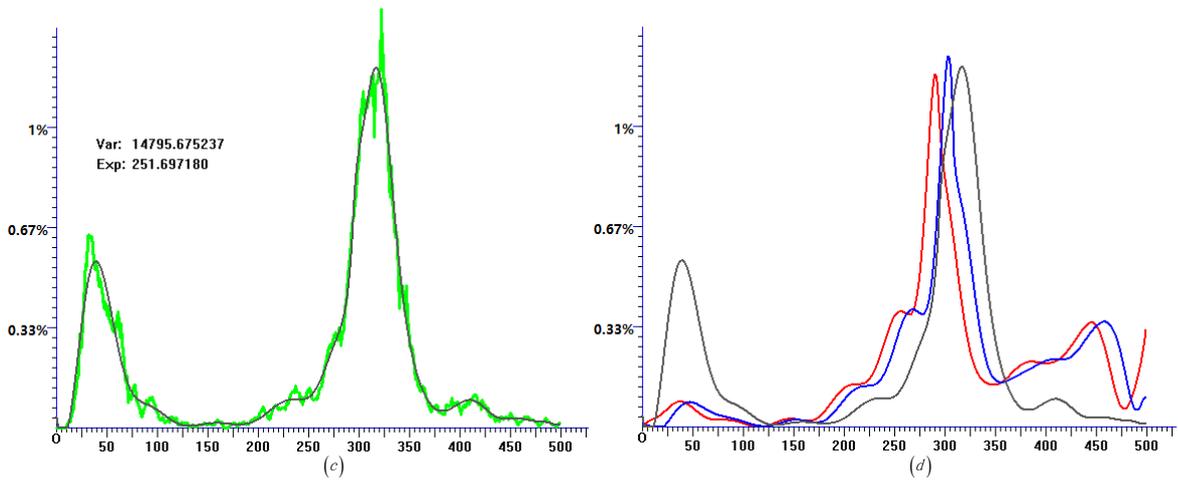

**Fig. 6.** Histogram data and quasi-distribution fitting of France

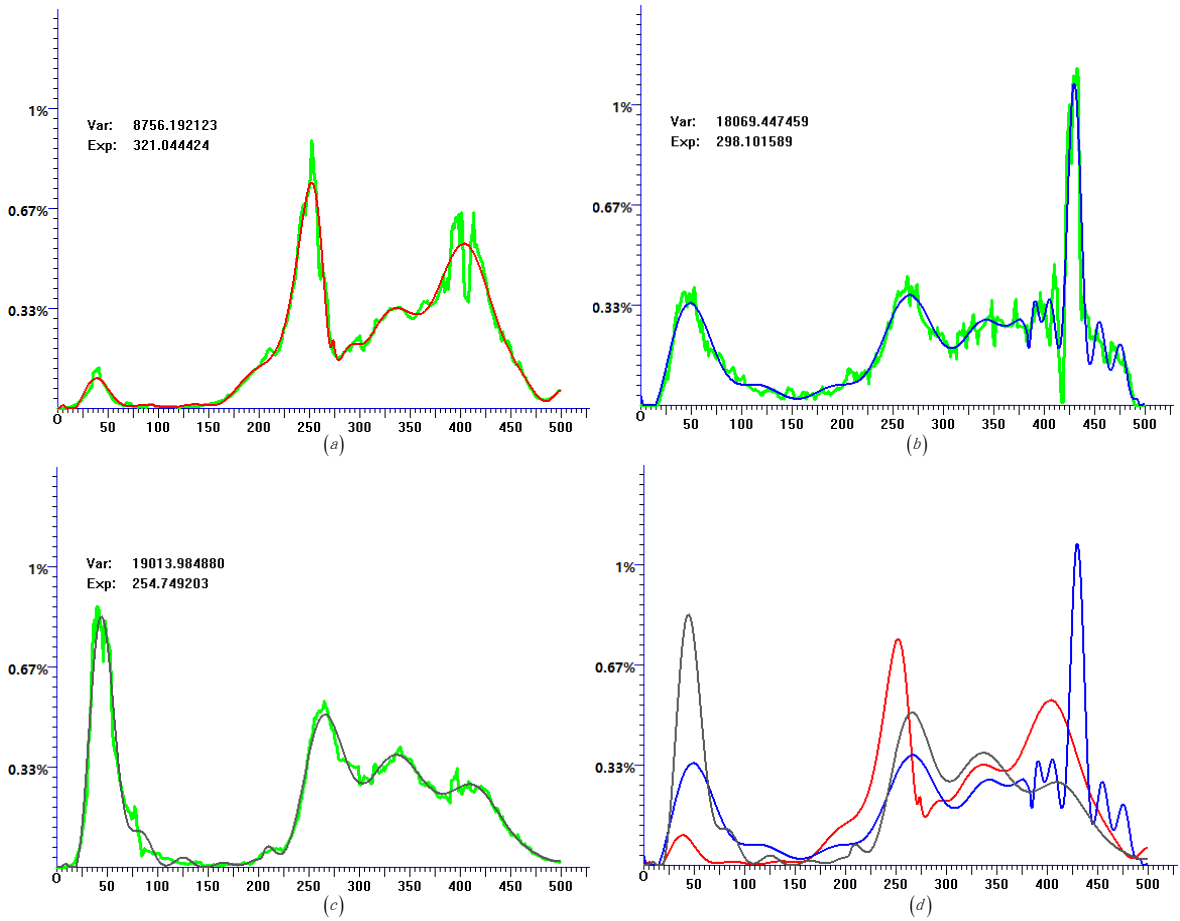

**Fig. 7.** Histogram data and quasi-distribution fitting of German

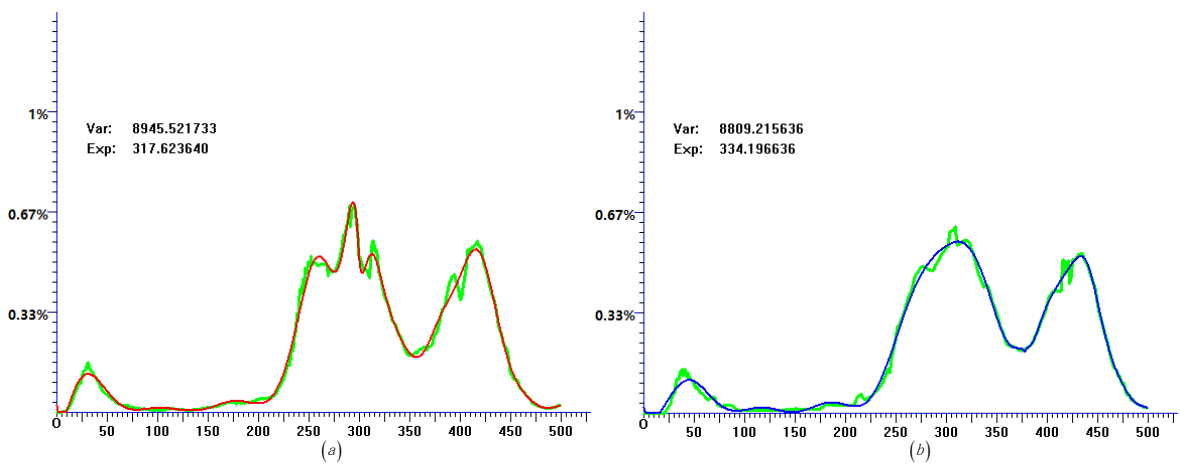

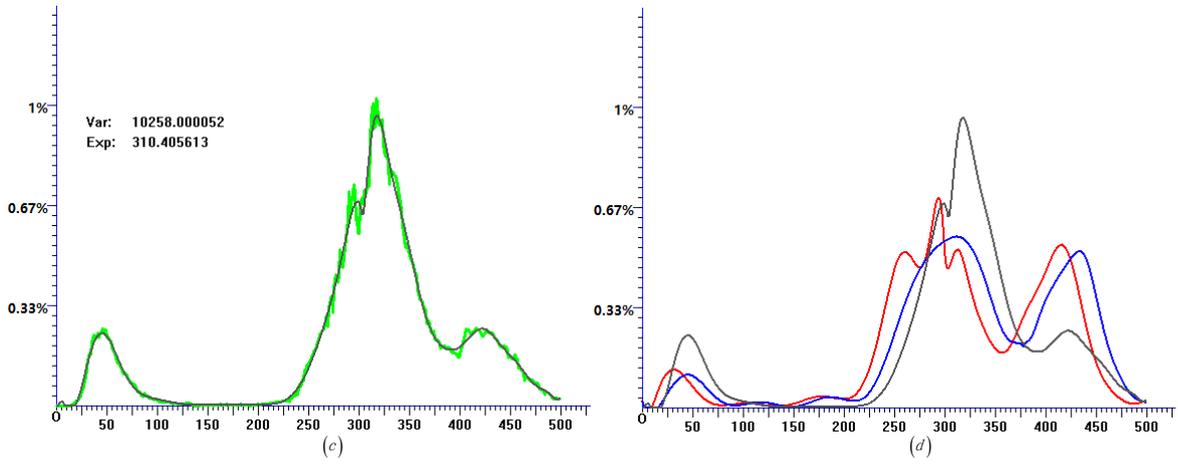

**Fig. 8.** Histogram data and quasi-distribution fitting of Italy

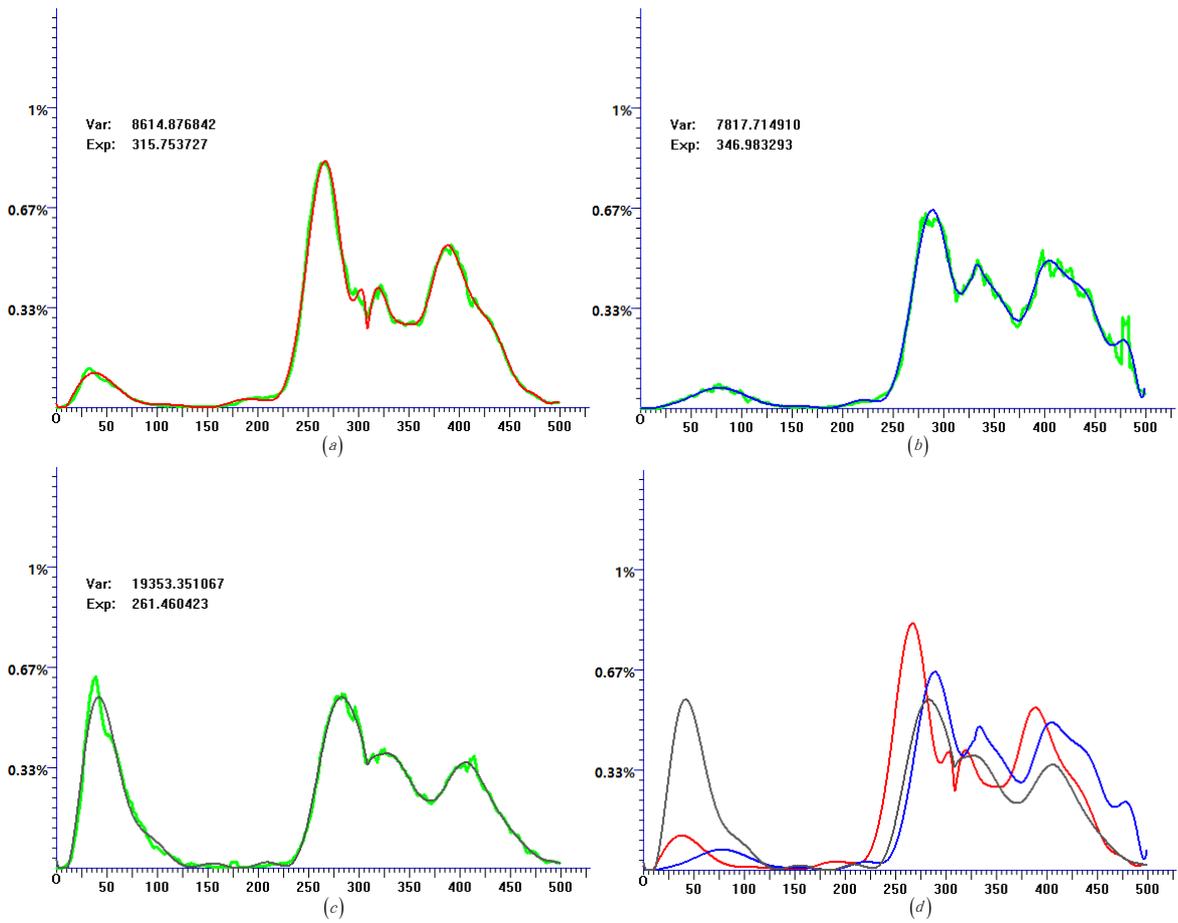

**Fig. 9.** Histogram data and quasi-distribution fitting of Iran

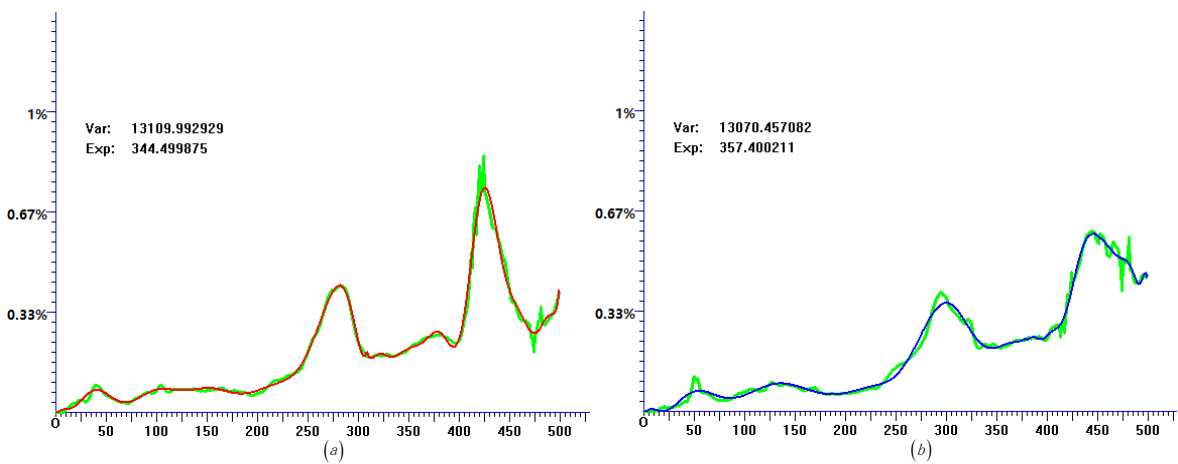

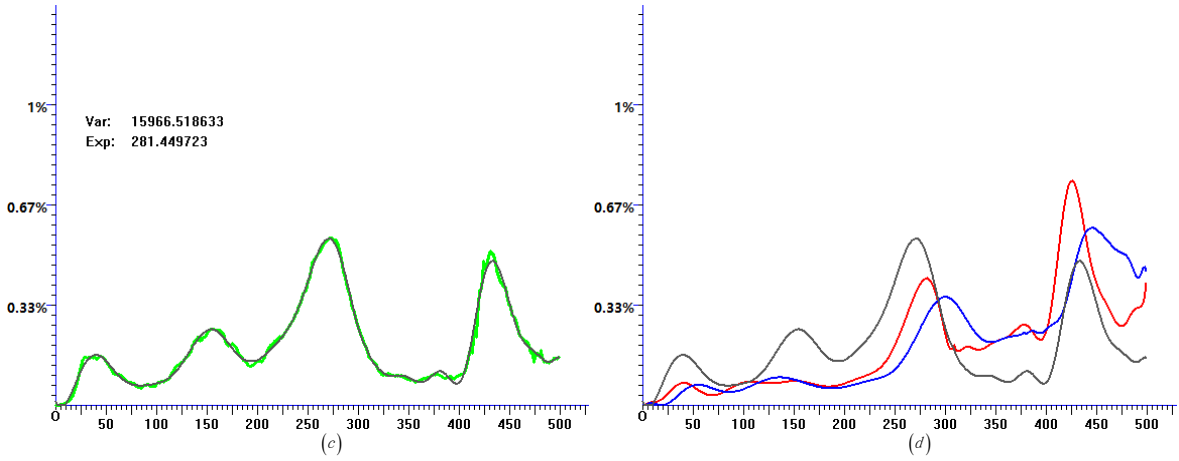

**Fig. 10.** Histogram data and quasi-distribution fitting of US

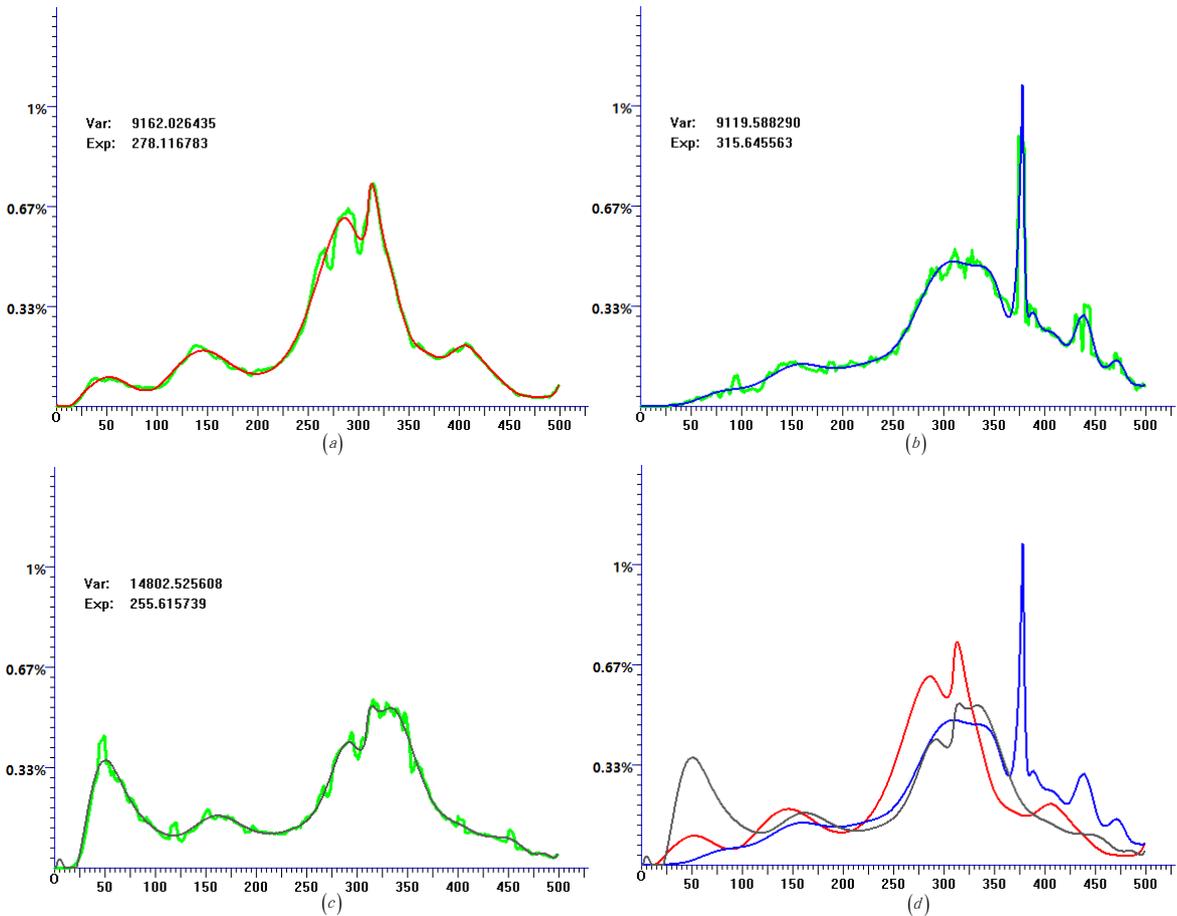

Figure 11 shows the fitting result of India, from (a), (b), (c) of the figure 10, we can see no matter daily confirmed cases, daily recovery cases or daily fatality cases, the histogram data and their corresponding quasi-distribution fitting curves both only have two peaks, it not means the COVID-19 rarely spread during the early 50~70 days, it just because the data around day 200 and day 430 (about the two peak points of abscissa) are too large to make the early days' data obviously. In figure 11 (d) we can find out the second peak of quasi-distribution fitting curve of daily fatality is lower than that of daily confirmed cases, it shows the declining of fatality rate of the pandemic. The same is true in Poland, show in figure 12.

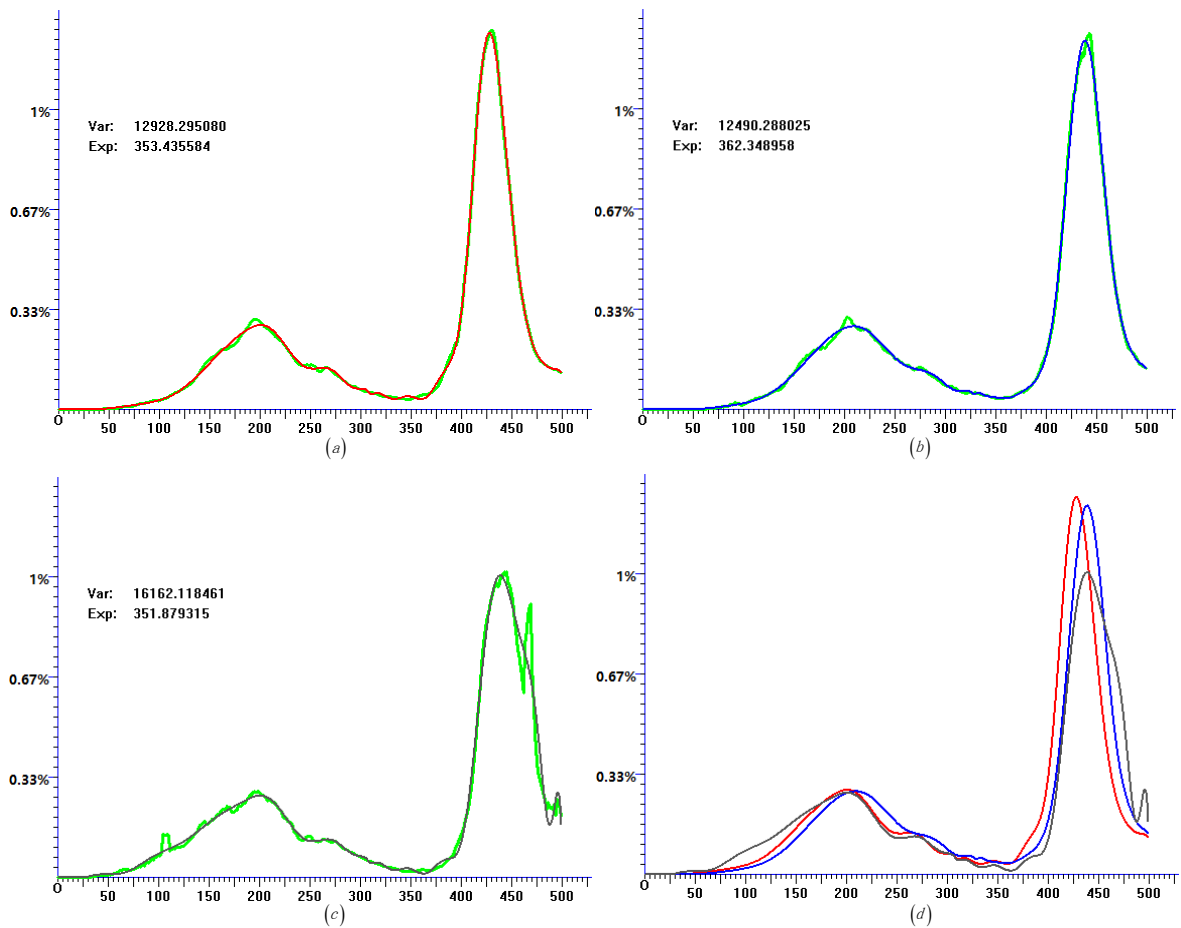

**Fig. 11.** Histogram data and quasi-distribution fitting of India

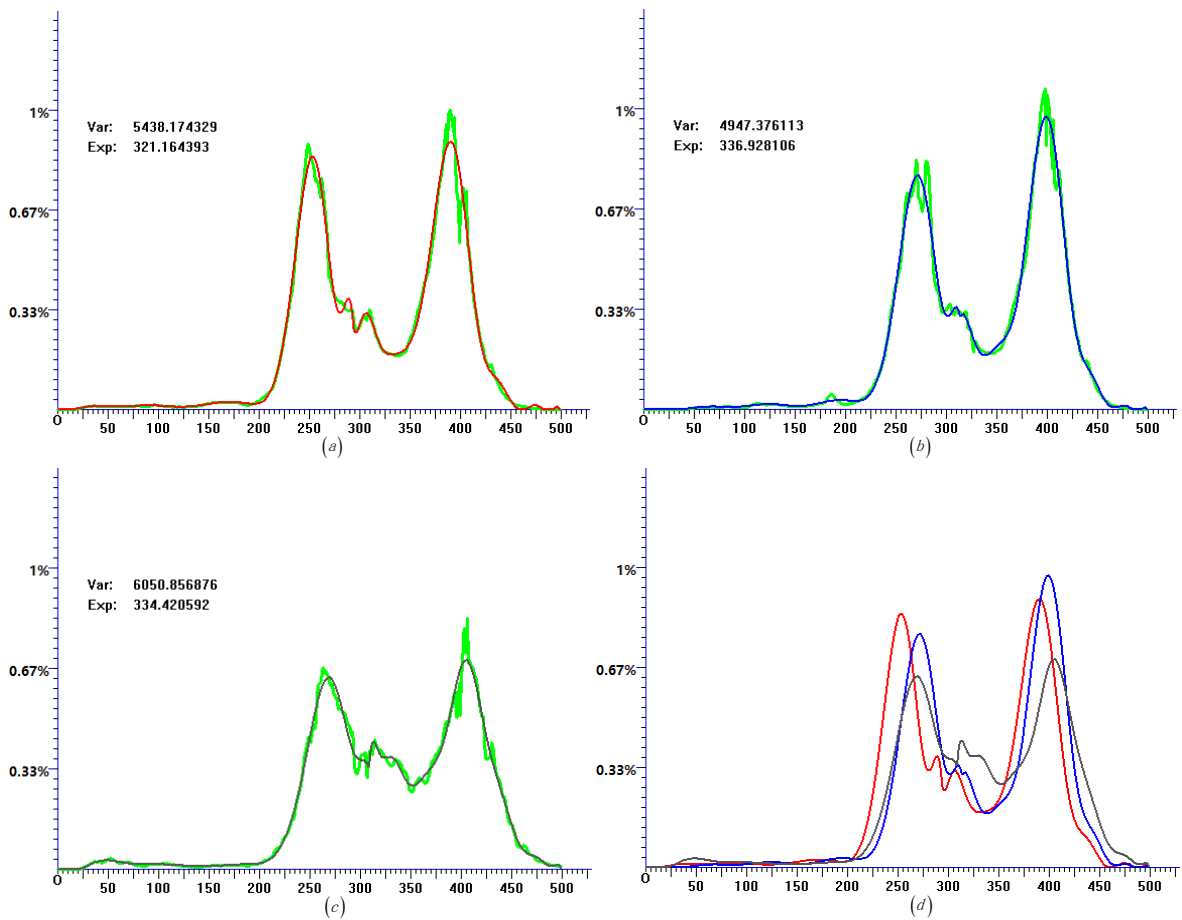

**Fig. 12.** Histogram data and quasi-distribution fitting of Poland

Figure 13 shows the fitting result of Israel, based on figure 13(d) even though the last peak of quasi-distribution fitting curve of daily fatality is lower than that of daily confirmed cases, but not very obviously, so does the peak before the last one, therefore, we should discreetly draw a conclusion that with the time going on the mortality rate has not increased in Israel. The epidemic situation in Japan and Philippines are similar, show in figure 14 and figure 15 respectively.

**Fig. 13.** Histogram data and quasi-distribution fitting of Israel

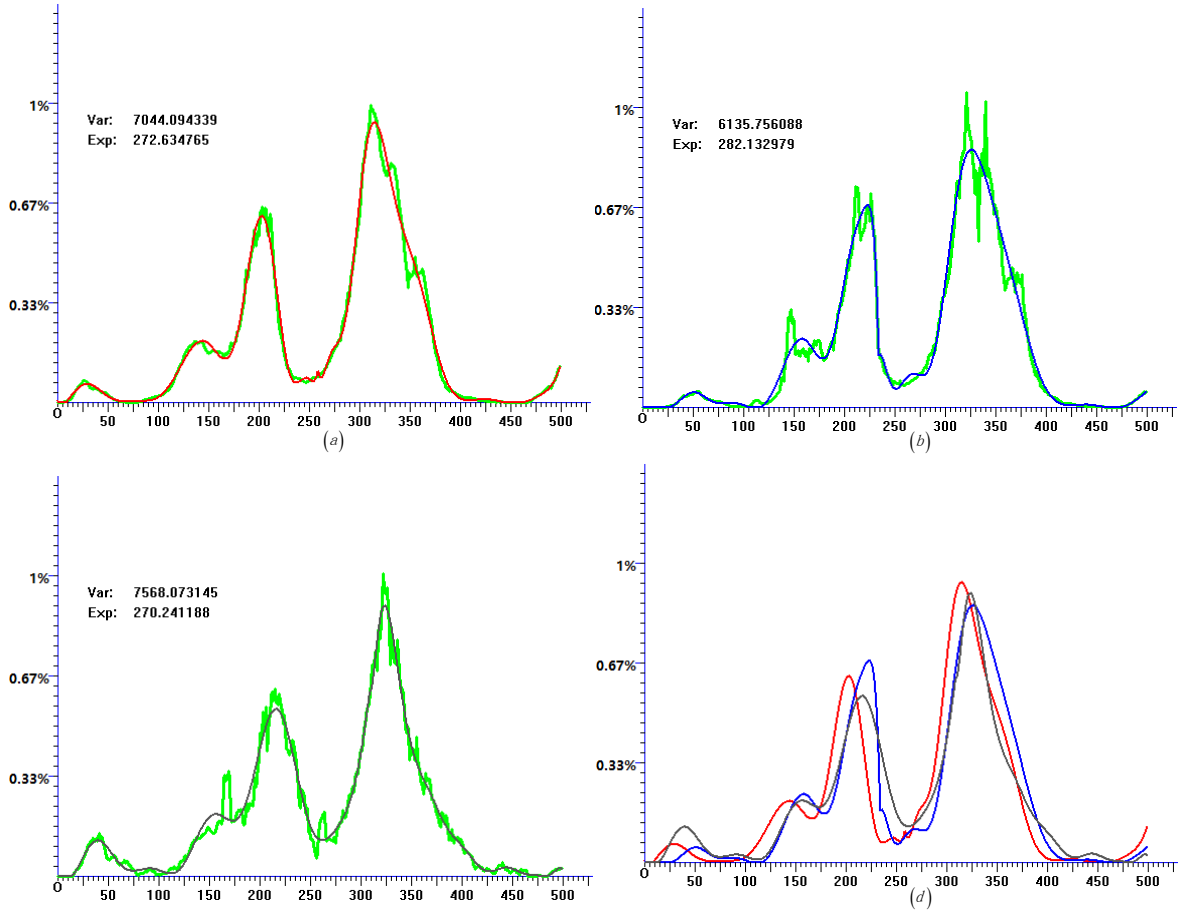

**Fig. 14.** Histogram data and quasi-distribution fitting of Japan

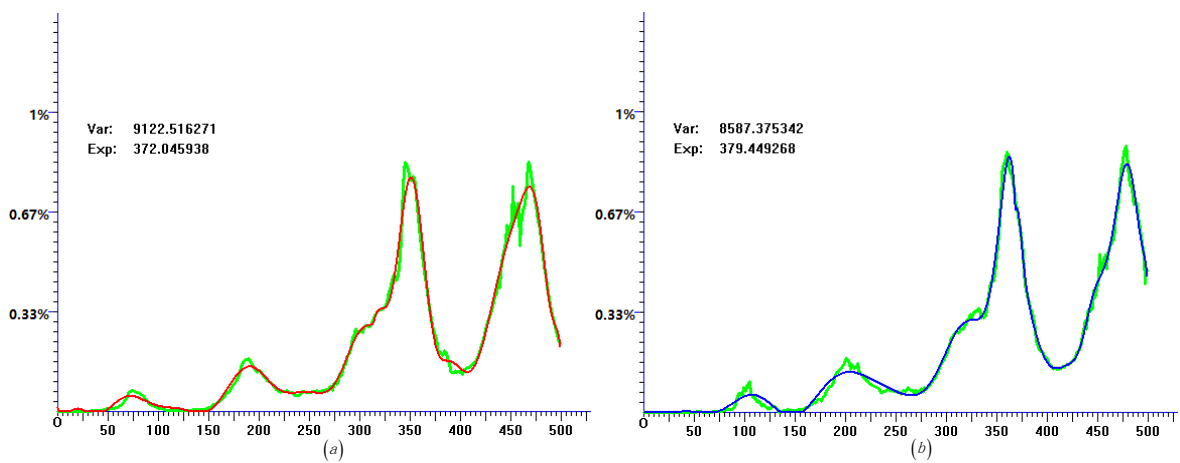

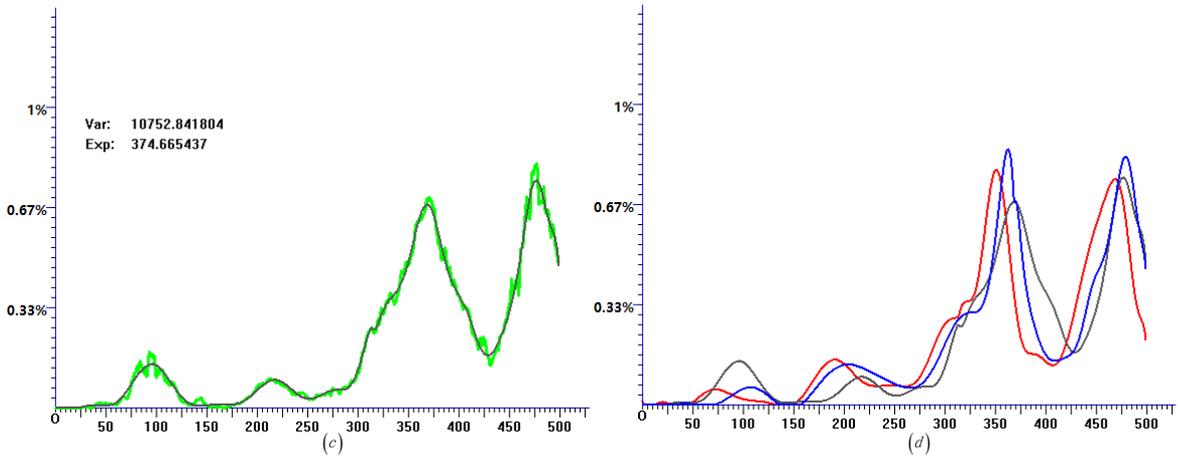

**Fig. 15.** Histogram data and quasi-distribution fitting of Philippines

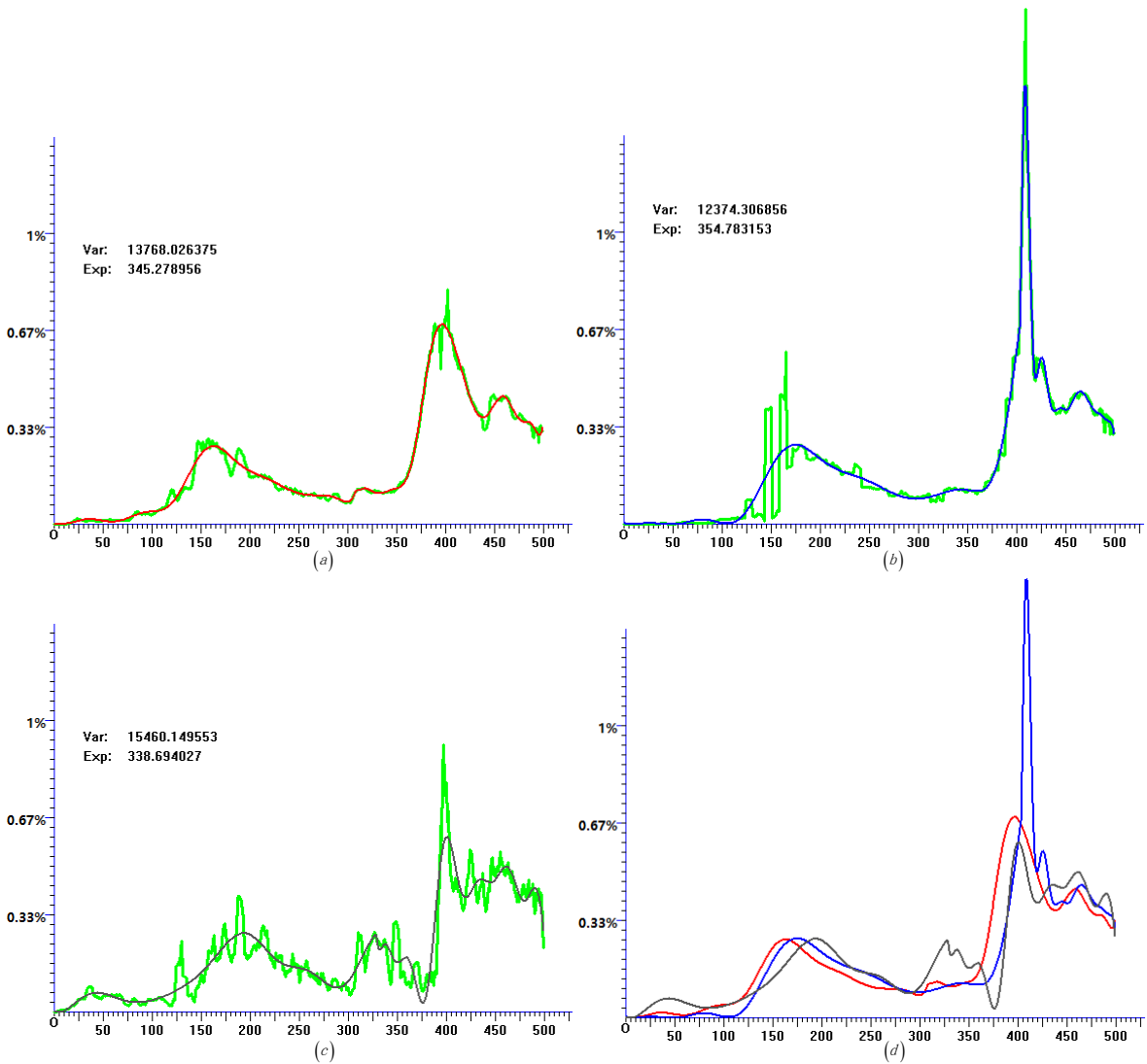

Figure 16 shows the fitting result of Korea, based on figure 16 (d), the penultimate peak of quasi-distribution fitting curve of daily fatality is much higher than the other two fitting curves, but at the last peak, the quasi-distribution fitting curve of daily fatality is obviously lower, which means the case fatality rate is declining at last. The epidemic situation in Lebanon and Portugal are similar, show in figure 17 and figure 18 respectively.

**Fig. 16.** Histogram data and quasi-distribution fitting of Korea

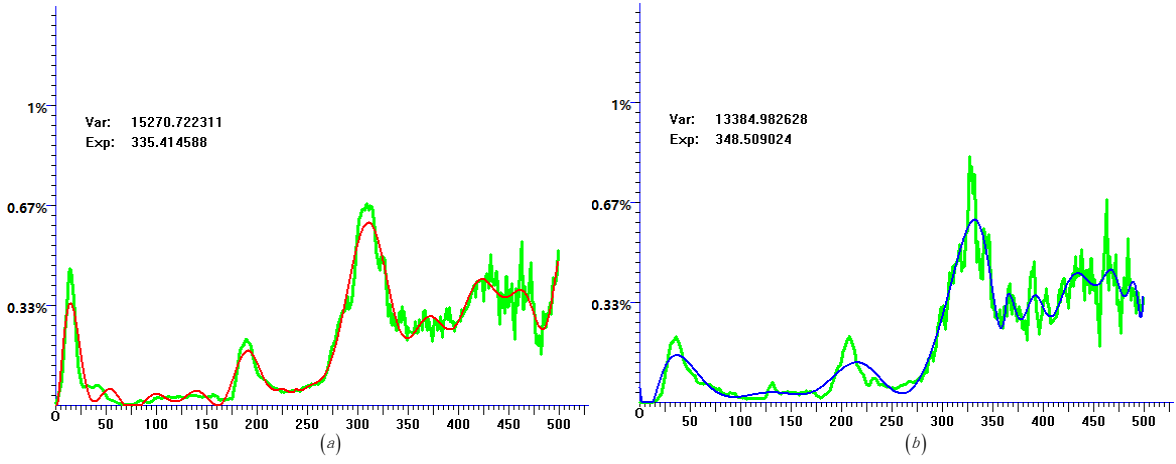
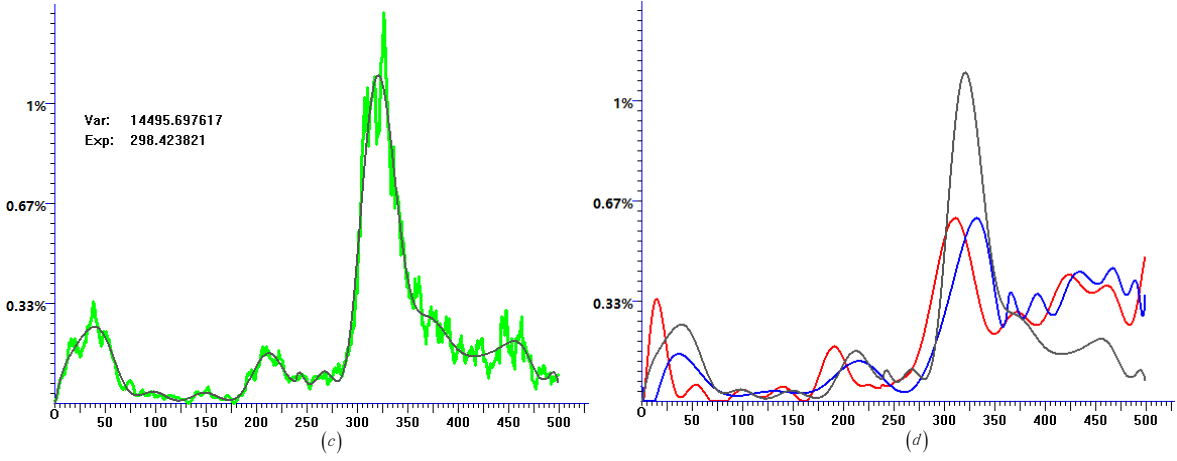

**Fig. 17.** Histogram data and quasi-distribution fitting of Lebanon

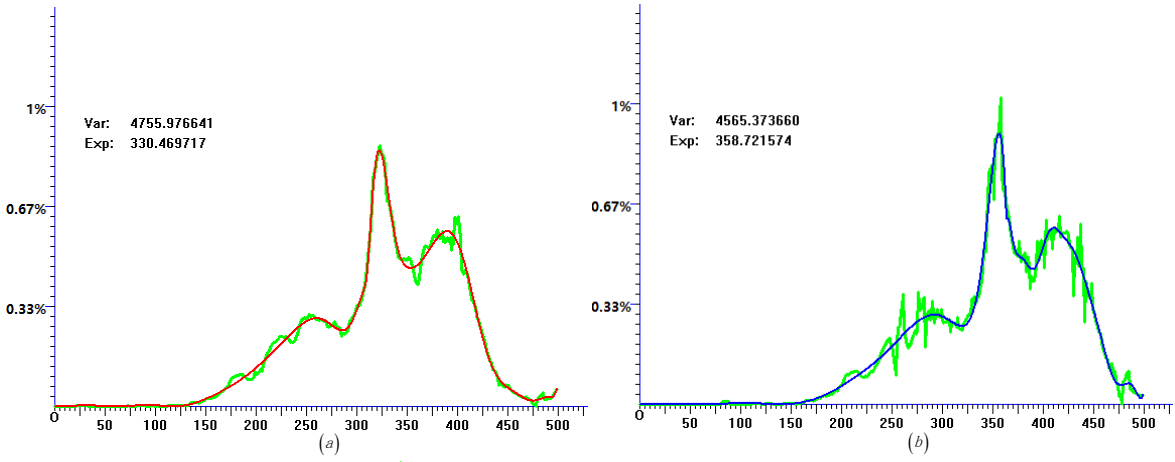
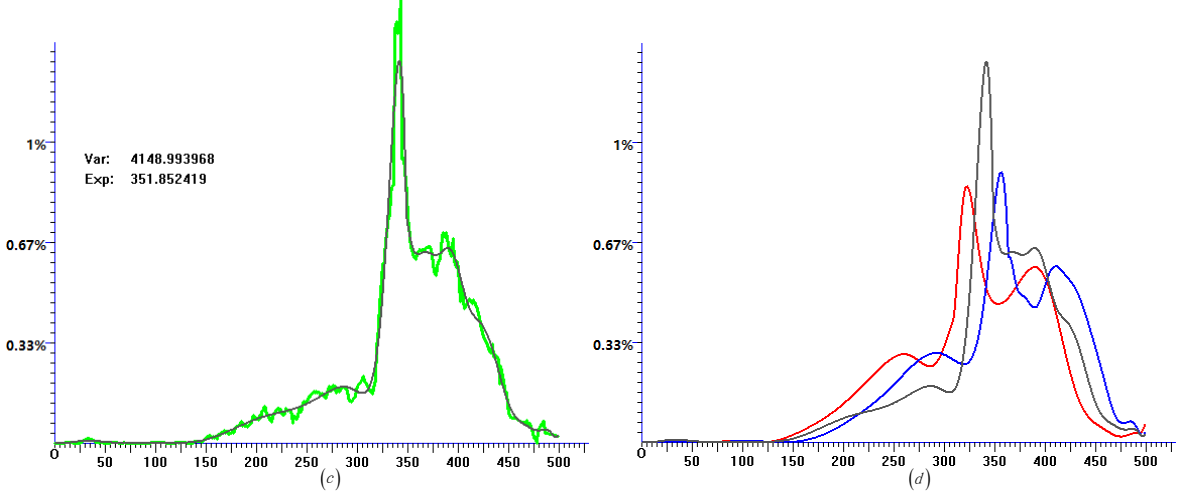

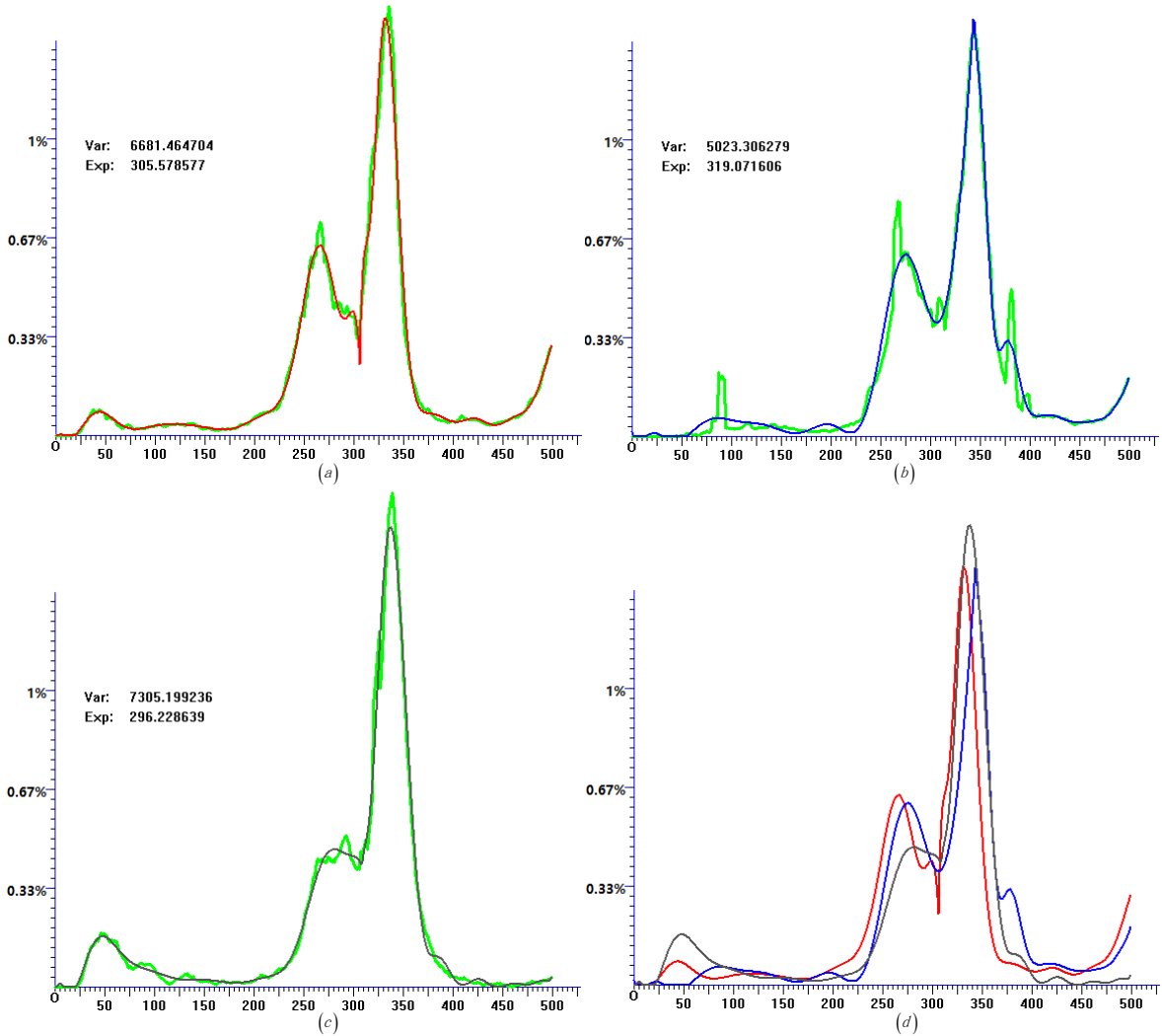

**Fig. 18.** Histogram data and quasi-distribution fitting of Portugal

## 4 Conclusion

In this paper, we build a new way, called QDF method to evaluate the spread of the epidemic caused by COVID-19 based on piecewise quasi-uniform B-spline curves. By fitting the distribution histogram data made from the daily confirmed cases, daily recovery cases and daily fatality cases of eighteen countries, we come to a conclusion that with the spread of the epidemic, even in the situation of the virus mutation, the case fatality rate of the COVID-19 is keep on declining. From the fitting results, the shape of the fitting curves of daily recovery cases, are much similar to the fitting curves of daily confirmed cases, showed in figure 1 (a), (b), figure 2 (a), (b), figure 3 (a), (b), figure 5 (a), (b), figure 7 (a), (b), figure 8 (a), (b), figure 9 (a), (b), figure 11(a), (b), figure 12 (a), (b), figure 13 (a), (b), figure 14 (a), (b), figure 16 (a), (b), figure 17 (a), (b), figure 18 (a), (b), and the corresponding fitting curves of daily recovery cases just certain days delay comparing with their daily confirmed cases fitting curves, which means the treatment of the epidemic caused by COVID-19 has stabilized, and in the near future, which is quite soon, the world will reopen.

**Acknowledgement**
We would like to thank the website of Toutiao and QQ, two of the biggest news website in China supply the data of the pandemic, their professional employee collected the epidemic data of almost every country from WHO, since outbreak of the epidemic.

**Appendix**

Assuming $T_i = t - i \times 0.1 \quad i = 0, 1, \cdots, 10$, then:

$$\tilde{N}_0 = \begin{cases} -10^5 T_1 & t \in [0, 0.1) \\ 0 & \text{otherwise} \end{cases}$$

$$\tilde{N}_1 = \begin{cases} 10^4 \left( 10 T_0 T_1^4 + 5 T_0 T_1^3 T_2 + \frac{5}{2} T_0 T_1^2 T_2^2 + \frac{5}{4} T_0 T_1 T_2^3 + \frac{5}{8} T_0 T_2^4 \right) & t \in [0, 0.1) \\ -6250 T_2^5 & t \in [0.1, 0.2) \\ 0 & \text{otherwise} \end{cases}$$

$$\tilde{N}_2 = \begin{cases} -10^4 \left( 5 T_0^2 T_1^3 + \frac{5}{2} T_0^2 T_1^2 T_2 + \frac{5}{4} T_0^2 T_1 T_2^2 + \frac{5}{8} T_0^2 T_2^3 + \frac{5}{3} T_0^2 T_1^2 T_3 + \frac{5}{6} T_0^2 T_1 T_2 T_3 + \frac{5}{12} T_0^2 T_2^2 T_3 + \frac{5}{9} T_0^2 T_1 T_3^2 + \frac{5}{18} T_0^2 T_2 T_3^2 + \frac{5}{27} T_0^2 T_3^3 \right) & t \in [0, 0.1) \\ 10^4 \left( \frac{5}{8} T_0 T_2^4 + \frac{5}{12} T_0 T_2^3 T_3 + \frac{5}{18} T_0 T_2^2 T_3^2 + \frac{5}{27} T_0 T_2 T_3^3 + \frac{5}{27} T_1 T_3^4 \right) & t \in [0.1, 0.2) \\ -\frac{50000}{27} T_3^5 & t \in [0.2, 0.3) \\ 0 & \text{otherwise} \end{cases}$$

$$\tilde{N}_3 = \begin{cases} 10^4\left(\frac{5}{3}T_0^3T_1^2 + \frac{5}{6}T_0^3T_1T_2 + \frac{5}{12}T_0^3T_2^2 + \frac{5}{9}T_0^3T_1T_3 + \frac{5}{18}T_0^3T_2T_3 + \frac{5}{27}T_0^3T_3^2 + \frac{5}{12}T_0^3T_1T_4 + \frac{5}{24}T_0^3T_2T_4 + \frac{5}{36}T_0^3T_3T_4 + \frac{5}{48}T_0^3T_4^2\right) & t\in[0,0.1) \\ -10^4\left(\frac{5}{12}T_0^2T_2^3 + \frac{5}{18}T_0^2T_2^2T_3 + \frac{5}{27}T_0^2T_2T_3^2 + \frac{5}{27}T_0T_1T_3^3 + \frac{5}{24}T_0^2T_2^2T_4 + \frac{5}{36}T_0^2T_2T_3T_4 + \frac{5}{36}T_0T_1T_3^2T_4 + \frac{5}{48}T_0^2T_2T_4^2 + \frac{5}{48}T_0T_1T_3T_4^2 + \frac{5}{48}T_1^2T_4^3\right) & t\in[0.1,0.2) \\ 10^4\left(\frac{5}{27}T_0T_3^4 + \frac{5}{36}T_0T_3^3T_4 + \frac{5}{48}T_0T_3^2T_4^2 + \frac{5}{48}T_1T_3T_4^3 + \frac{5}{48}T_2T_4^4\right) & t\in[0.2,0.3) \\ -\frac{3125}{3}T_4^5 & t\in[0.3,0.4) \\ 0 & \text{otherwise} \end{cases}$$

$$\tilde{N}_4 = \begin{cases} -10^4\left(\frac{5}{12}T_0^4T_1 + \frac{5}{24}T_0^4T_2 + \frac{5}{36}T_0^4T_3 + \frac{5}{48}T_0^4T_4 + \frac{1}{12}T_0^4T_5\right) & t\in[0,0.1) \\ 10^4\left(\frac{5}{24}T_0^3T_2^2 + \frac{5}{36}T_0^3T_2T_3 + \frac{5}{36}T_0^2T_1T_3^2 + \frac{5}{48}T_0^3T_2T_4 + \frac{5}{48}T_0^2T_1T_3T_4 + \frac{5}{48}T_0T_1^2T_4^2 + \frac{1}{12}T_0^3T_2T_5 + \frac{1}{12}T_0^2T_1T_3T_5 + \frac{1}{12}T_0T_1^2T_4T_5 + \frac{1}{12}T_1^3T_5^2\right) & t\in[0.1,0.2) \\ -10^4\left(\frac{5}{36}T_0^2T_3^3 + \frac{5}{48}T_0^2T_3^2T_4 + \frac{5}{48}T_0T_1T_3T_4^2 + \frac{5}{48}T_0T_2T_4^3 + \frac{1}{12}T_0^2T_3^2T_5 + \frac{1}{12}T_0T_1T_3T_4T_5 + \frac{1}{12}T_0T_2T_4^2T_5 + \frac{1}{12}T_1^2T_3T_5^2 + \frac{1}{12}T_1T_2T_4T_5^2 + \frac{1}{12}T_2^2T_5^3\right) & t\in[0.2,0.3) \\ 10^4\left(\frac{5}{48}T_0T_4^4 + \frac{1}{12}T_0T_4^3T_5 + \frac{1}{12}T_1T_4^2T_5^2 + \frac{1}{12}T_2T_4T_5^3 + \frac{1}{12}T_3T_5^4\right) & t\in[0.3,0.4) \\ -\frac{2500}{3}T_5^5 & t\in[0.4,0.5) \\ 0 & \text{otherwise} \end{cases}$$

$$\tilde{N}_5 = \begin{cases} \frac{2500}{3}T_0^5 & t\in[0,0.1) \\ -\frac{2500}{3}\left(T_0^4T_2 + T_0^3T_1T_3 + T_0^2T_1^2T_4 + T_0T_1^3T_5 + T_1^4T_6\right) & t\in[0.1,0.2) \\ \frac{2500}{3}\left(T_0^3T_3^2 + T_0^2T_1T_3T_4 + T_0^2T_2T_4^2 + T_0T_1^2T_3T_5 + T_0T_1T_2T_4T_5 + T_0T_2^2T_5^2 + T_1^3T_3T_6 + T_1^2T_2T_4T_6 + T_1T_2^2T_5T_6 + T_2^3T_6^2\right) & t\in[0.2,0.3) \\ -\frac{2500}{3}\left(T_0^2T_4^3 + T_0T_1T_4^2T_5 + T_0T_2T_4T_5^2 + T_0T_3T_5^3 + T_1^2T_4^2T_6 + T_1T_2T_4T_5T_6 + T_1T_3T_5^2T_6 + T_2^2T_4T_6^2 + T_2T_3T_5T_6^2 + T_3^2T_6^3\right) & t\in[0.3,0.4) \\ \frac{2500}{3}\left(T_0T_5^4 + T_1T_5^3T_6 + T_2T_5^2T_6^2 + T_3T_5T_6^3 + T_4T_6^4\right) & t\in[0.4,0.5) \\ -\frac{2500}{3}T_6^5 & t\in[0.5,0.6) \\ 0 & \text{otherwise} \end{cases}$$

$\tilde{N}_6(t) = \tilde{N}_5(t-0.1)$ , $\quad \tilde{N}_7(t) = \tilde{N}_5(t-0.2)$ , $\quad \tilde{N}_8(t) = \tilde{N}_5(t-0.3)$ , $\quad \tilde{N}_9(t) = \tilde{N}_5(t-0.4)$ ,

$\tilde{N}_{10}(t) = \tilde{N}_4(1-t)$ , $\quad \tilde{N}_{11}(t) = \tilde{N}_3(1-t)$ , $\quad \tilde{N}_{12}(t) = \tilde{N}_2(1-t)$ , $\quad \tilde{N}_{13}(t) = \tilde{N}_1(1-t)$ ,

$\tilde{N}_{14}(t) = \tilde{N}_0(1-t)$ 。